\documentclass[12pt,english]{article}
\usepackage[T1]{fontenc}
\usepackage[latin1]{inputenc}  
\usepackage{geometry}
\usepackage{amsmath, amsthm, amssymb}
\usepackage{verbatim}
\usepackage{babel}
\usepackage{url}
\usepackage{graphics}

\makeatletter 

\renewcommand{\baselinestretch}{1.5}
\newcommand{\slbf}[1]{\textbf{#1}}

\newcounter{claimcounter}[section] 
\renewcommand{\theclaimcounter}{\arabic{section}.\arabic{claimcounter}}
\newenvironment{theorem}{\begin{flushleft}\refstepcounter{claimcounter}\slbf{Theorem
      \theclaimcounter{} }\sl}{\end{flushleft}}

\newenvironment{claim}{\begin{flushleft}\refstepcounter{claimcounter}\slbf{Claim \theclaimcounter{} }\sl}{\end{flushleft}}
\newenvironment{corollary}{\begin{flushleft}\refstepcounter{claimcounter}\slbf{Corollary
        \theclaimcounter{} }\sl}{\end{flushleft}}

\newenvironment{atheorem}[1]{\begin{flushleft}\slbf{Theorem #1 }\sl}{\end{flushleft}}

\newcommand{\WP}{$W$-problem }
\newcommand{\SP}{$S$-problem }
\newcommand{\K}{\mathcal{K}}

\newcommand{\half}{\frac{1}{2}}

\renewcommand{\proof}{\textbf{Proof. }}

\newcommand{\p}{^{\prime}}
\newcommand{\pp}{^{\prime\prime}}
\newcommand{\Gx}{G_{\mathcal{X}}}
\newcommand{\kx}{\mathcal{K}_{\mathcal{X}}}

\newcommand{\x}{\mathcal{X}}
\newcommand{\y}{\mathcal{Y}}

\newcommand{\etax}{\eta_{\x}}

\newcommand{\thetax}{\theta_{\x}}
\newcommand{\thetay}{\theta_{\y}}

\newcommand{\GTK}{(G,T,\K)}
\newcommand{\GTKp}{(G\p,T\p,\K \p)}
\newcommand{\GTKpp}{(G\pp,T\pp,\K \pp)}

\newcommand{\GTKx}{(\Gx,\x,\kx)}

\newcommand{\E}{\mathcal{E}}
\newcommand{\F}{\mathcal{F}}

\newcommand{\Lovasz}{Lov\~{a}sz}
\newcommand{\quarter}{\frac{1}{4}}
\newcommand{\GT}{(G,(T,S))}
\newcommand{\heta}{\hat{\eta}}

\date{}
\makeatother
\begin{document}
\title{On fractionality of the path packing problem}
\author{Natalia Vanetik\footnote{Department of Computer Science, Ben-Gurion University, Israel, \textit{orlovn@cs.bgu.ac.il}}}

\maketitle
\vspace{-1cm}

\begin{abstract}
\small{In an undirected graph $G$ with node set $N$ and a subset $T\subseteq N$,
  a \textit{fractional multiflow problem} is defined as finding 
  $\max_{f} \sum_{(u,v)}\omega(u,v)f[u,v]$ over all collections $f$ of
  weighted paths with ends in $T$ (the $\omega$\textit{-problem}).
  $f[u,v]$ denotes the total weight
  of paths with the end-pair $(u,v)$ in $f$.
  The paths of $f$ must
  satisfy the edge capacity constraint: total weight of
  the paths traversing a single edge does not exceed $1$.
  We study a fractional multiflow problem with the reward function $\omega$ having values
  $(0, 1)$ (a \textit{fractional path packing problem}),
  and an auxiliary \textit{weak problem} where $\omega$ is a metric.
  A. Karzanov in \cite{k89} defined the {\it fractionality}
  of $\omega$ with respect to a 
  given class of networks $(G,T)$ as the least natural $D$ such that
  for any network $(G,T)$ from the class, the $\omega$-problem has a
  solution which becomes integer-valued when multiplied by $D$.
  He proved that a fractional path packing problem has infinite
  fractionality outside a very specific class of networks, and conjectured
  that within this class, the fractionality does not exceed $4$
  ($2$ for Eulerian networks).
  In this paper we prove Karzanov's conjecture by showing that the
  fractionality of both fractional path packing and weak problems
  is $1$ or $2$ for every Eulerian network in this class.} 
\end{abstract}
\section{Introduction}
In this paper we study collections of edge-disjoint paths in a network, 
also called \textit{paths packings} or \textit{multiflows}, addressing an optimization problem
of  the following form. Let $G = (N, E)$ be a multigraph with node-set $N$
and edge-set $E$, and let $T\subseteq N$ be a set of nodes distinguished as \textit{terminals}. 
By a $T${\it -path} we mean an unclosed path with the ends in T, and by an
{\it integer $T$-flow}, or an integer multiflow,
we mean a collection of pairwise edge-disjoint $T$-paths in $G$. 
Let us define a \textit{fractional $T$-flow} as a non-negative weight function $f (P )$
on the set of all $T$-paths in $(G, T )$, satisfying the {\it edge capacity}
constraints:
\begin{equation}\label{capacity-c}\mbox{\sl $\sum_{P}f(P)I(P,(x,y))\leq
c(x,y)$ for each adjacent pair $(x, y)$ of nodes in
$N$}\end{equation}
Here $I(P, (x, y))$ denotes the number of $(x,y)$-edges of
$G$ traversed by $P$,
and $c(x, y)$ is the edge capacity, equal to the number of
$(x,y)$-edges in $G$.
Given non-negative "rewards" $\omega(u, v)$ assigned to the unordered pairs of
terminals, the problem is to
\begin{equation}\label{omega-problem}\mbox{\sl maximize
    $\sum_{u,v}\omega(u,v)f[u,v]$ over the fractional $T$-flows $f$ in $(G,
    T )$,}\end{equation}
where $f[u,v]$  denotes the total weight of the $(u,v)$-paths in $f$.
For short, \eqref{omega-problem} will be referred to as the
$\omega$\textit{-problem}. 
This is one of the basic multiflow problems, 
having numerous applications, such as communication and VLSI design.
Not surprisingly, for most reward functions the $w$-problem is known to be
$\mathbb{NP}$-hard over integer multiflows,
not only when a network $(G, T )$ is quite arbitrary, 
but even for such friendly classes as the planar or the 
Eulerian networks (the latter class is studied in this paper).

However, the more fragmented is $f$ between various paths, the less is its
utility 
for discrete path packing. 
To make this precise, let us, following A. Karzanov \cite{k89}, 
define the {\it fractionality} of the reward function $\omega$ with respect to a 
given class of networks $(G, T )$: this is the least natural $D$ such that
for any network $(G, T )$ from the class, the $\omega$-problem has a
solution $f$ which becomes integer-valued when multiplied by $D$ 
(in short, a $\frac{1}{D}$ -integer solution). 
For certain reward functions, fractionality for the general networks was
found to be $2$ (see \cite{ikl00} and \cite{l04}); for some of them, the
$\omega$-problem 
was also shown to have an integer solution provided that the non-terminal
    (\textit{inner}) nodes of
a network have even degrees; such networks are called \textit{Eulerian}.

Two specific classes of the reward function 
are of principal importance. One comprises the $(0, 1)$ reward functions. 
It is convenient to represent such a function by a demand graph (or scheme) 
$(T, S)$ where $S: = {(u, v):\:  \omega(u, v) = 1}$, 
and to call \eqref{omega-problem} the {\it $S$-problem}. 
Let a path in $G$ be called an \textit{$S$-path} if its end-pair belongs to $S$, and 
a collection of $S$-paths satisfying \eqref{capacity-c} be called an $S$-flow. 
Thus, the \SP may be stated as maximizing of \\$f [S]: = \sum_{(u,v)\in S}f[u,v]$.
A. Karzanov has described the
fractionality of the $(0, 1)$ reward functions (or the schemes $S$) in \cite{k89}. Namely,
the fractionality of $S$ is finite iff 
any distinct pairwise intersecting anticliques 
(i.e., inclusion-maximal stable sets) $A, B, C$ of $(T, S)$ satisfy
\begin{equation}\label{kc}A\cap B = A\cap C = B\cap C,\end{equation} 
and the finite fractionality can only equal $1$, $2$, or $4$.
He conjectured that this
\begin{equation}\label{k-conjecture}\mbox{\sl finite fractionality can only be $1$ or $2$.}\end{equation} 
%We denote the network $(G,T)$ together with the anticlique clutter
%$\K$ of $(T, S)$ by $\GTK$.

Not long ago, H. Ilani and E. Barsky observed that the problem of discrete path
packing is $\mathbb{NP}$-hard, even for Eulerian networks, 
for each demand graph violating \eqref{kc}. 
So, investigating the \SP has focused on the schemes satisfying
\eqref{kc}.
In this paper we consider the \SP for $S$ satisfying \eqref{kc}
together with an auxiliary weak problem, denoted a $W$\textit{-problem}:
an $\omega$-problem where $\omega$ 
is a metric defined by $\omega(u, v) = 1$ for $(u, v)\in S$, $\half$ for
$(u, v)$ covered by
exactly one anticlique of $(T, S)$, and $0$ for the others 
(i. e., those covered by at least two anticliques). An anticlique clutter of $(T,S)$
satisfying \eqref{kc}
is called a \textit{K-clutter}, and an Eulerian network $\GTK$ with an
anticlique K-clutter $\K$ of $(T,S)$
is called a \textit{K-network}.
The maxima of
$S$- and $W$-problems are denoted by $\eta$ and $\theta$ respectively.

In this paper, we prove conjecture \eqref{k-conjecture}.
Additionally, we show that the \WP in a K-network also admits a
solution of fractionality at most $2$. We use the following crucial fact:
every \SP and \WP in a network satisfying \eqref{kc}
have a common solution (Theorem 1 of \cite{va07a}).

\begin{figure}[!t]
{\par \centering
  \resizebox*{0.7\textwidth}{0.2\textheight}{\includegraphics{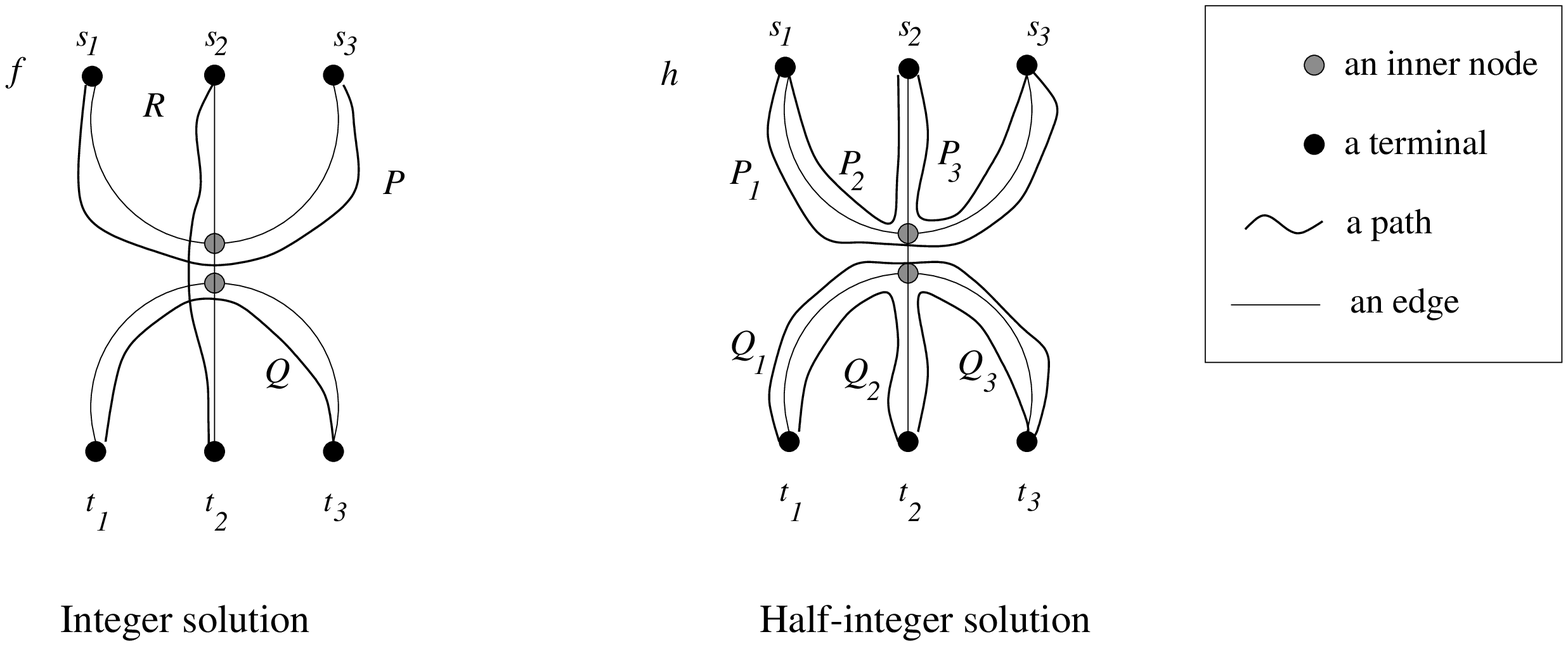}} \par}
\caption{\label{non-integral-figure}The fractionality of \SP and \WP can be $2$.}
\end{figure}

The bound on fractionality is tight in both cases, as an example in
Figure \ref{non-integral-figure} demonstrates.
There we have $\K=\{\{s_{i},t_{j}\}\}$,
$i,j\in \{1,2,3\}$, and every integer multiflow in this network has no more than
$2$ $S$-paths, for example, paths $P$ and $Q$ in Figure
\ref{non-integral-figure}(a). The maximum of the \WP among integer
multiflows is $2\half$. 
However, in this network there exists a half-integer multiflow
$h=\{P_{1},P_{2},P_{3},Q_{1},Q_{2},Q_{3}\}$
with weight of every path $\half$ being (see Figure
\ref{non-integral-figure}(b)).
The value of $\sum_{u,v}\omega(u,v)h[u,v]$ for both \SP and \WP is $3$.
Thus, an integer solution to the \SP or the \WP does not always exist. 
Table \ref{defs-table} summarizes notation used in this paper.    
\begin{table}[!h]
  \small{
  \begin{tabular}{|l|l|}
    \hline
    \textbf{Notation} & \textbf{Definition} \\ \hline \hline
    $\GTK$ & a network $\GT$ and the anticlique clutter $\K$ of $(T,S)$\\
    \hline
    $S$-path & a path whose end-pair is in $S$\\ \hline
    $W$-path & a path whose end-pair is covered by exactly one member of $\K$\\ \hline
    zero path & a path whose end-pair is covered by two members of $\K$\\ \hline
    $d(X)$, $X\subset N$ & the number of $(X,\overline{X})$-edges in $G$\\
    \hline
    $\lambda(A)$, $A\subseteq T$ & $\min\{d(X): \; X\subset N,\;\;X\cap
T=A\}$\\ \hline
    $\beta(A)$, $A\subseteq T$ & $\frac{1}{2}(\sum_{t\in A}\lambda
    (t)-\lambda(A))$; is an integer in Eulerian networks\\ \hline
    $A^{c}$, $A\subseteq T$ & $T\setminus A$\\ \hline
    $\overline{A}$, $A\subseteq N$ & $N\setminus A$\\ \hline
    an $(A,B)$-path (an $A$-path), $A,B\subseteq N$ & a path ends in $A$ and
    $B$ (in $A)$\\ \hline
    $f[A,B]$ & the number of $(A,B)$-paths in $f$ ($f[A]$ when
$A=B$) \\ \hline
    $w(P)$ & the weight of path $P$ \\ \hline
%    $I(P,e)$ & the number of times path $P$ traverses edge $e$ \\ \hline
%    $\sum_{P\in h}I(P,e)w(P)$ & the total weight of the paths of $h$ traversing
%    an edge $e$ \\ \hline
    $xPy$ & an $(x,y)$-segment of a path $P$, where $x$ and $y$ are nodes\\
    \hline
    $|f|$ & the size of a multiflow $f$: the total weight of its paths\\ \hline
    a maximum multiflow & a multiflow of maximum size \\ \hline
    the fractionality of a multiflow & the largest denominator among its paths'
    weights\\ \hline
%   a multiflow $f$ un/under/saturates $e\in E$ & $\sum_{P\in f}I(P,e)w(P)$ is
%    $0$/between $0$ and $1$/is $1$\\ \hline
%    $\phi(h)$ & $\sum_{t\in T}(\lambda(t)-h[t,t^{c}])$ for multiflow $h$\\
%    \hline
%    $\phi_{A}(h)$, $A\subseteq T$ & $\sum_{t\in A}(\lambda(t)-h[t,t^{c}])$ for multiflow $h$\\
%    \hline
    $s\sim t$, $s,t\in T$ & $(s,t)$ is a zero pair\\ \hline
    an atom & a set of terminals not separated by a member of $\K$\\ \hline
%    $\J$ & the set of all atoms\\ \hline
%    $\gamma_{J}$, $J\in \J$ ($\gamma_{t}$ if $J=\{t\}$)& number of $\K$-members
%    containing $J$ \\ \hline
    $\K$ is simple & every atom in $\K$ has size $1$\\ \hline
%    $n$-inflation $\nGTK{n}$ of $\GTK$ & multiplying $n$ times every
%edge from $E$\\ \hline
 %   $n$-inflation $nh$ of a flow $h$ & multiplying by $n$ the weight of
%    every path in $h$ \\ \hline 
  \end{tabular}}
  \caption{\label{defs-table}Notation}  
\end{table}

\section{Outline of the proof}
We observe K-networks that are counterexamples to the fractionality
conjecture for either $W$- or $S$-problem.
%If the fractionality
%is $3$, $2$-inflation $\nGTK{2}$ admits a half-integer multiflow with the
%required properties, and its $2$-compression is exactly the quarter-integer multiflow we
%seek.
First, we prove the fractionality conjecture for the $W$-problem
by showing that a half-integer simple multiflow of the smallest size solving
the $W$-problem exists.
Second, we observe a minimal K-network that fails to satisfy the \SP
    fractionality conjecture and show that it admits a half-integer solution.
\section{\label{locking-section}Operations on paths and locking}
A pair of paths with disjoint end-pairs and
a common node forms a \textit{cross}.
A path is \textit{compound} if it traverses a terminal different from its ends, and
\textit{simple} otherwise. A multiflow is called \textit{simple} if
it contains only simple paths.
 
Let paths $P$ and $Q$ of a multiflow $f$ traverse an inner node $x$, 
so that $P=P\p xP\pp$ and $Q=Q\p xQ\pp$. \textit{Switching} $P$ and $Q$ in $x$ 
transforms them into $K=P\p xQ\p$ and $L=P\pp xQ\pp$ and $f$ into the multiflow
$f\setminus\{P,Q\}\cup\{K,L\}$. 
A \textit{split} of an inner node $x$ is a graph transformation consisting
of removal of $x$ and linking its neighbors by $\frac{d(x)}{2}$ edges
so as to preserve their degrees. Given a multiflow $h$ in a network, 
an $h$\textit{-split} of an inner node is a split preserving the paths 
of $h$.
%\textit{Splitting-off} a pair of edges incident to a node $x$ consists of removing
%these edges and linking their non-$x$-ends by a single edge.

A maximum multiflow $f$ \textit{locks} a set $A\subseteq T$ if it contains a maximum 
$(A,A^{c})$-flow, that is, if $f[A,A^c]=\lambda(A)$.
Otherwise, $f$ \textit{unlocks} $A$. In other words, $f$ locks $A$ if it
contains the smallest possible number of $A$-paths.
A. Karzanov and M. Lomonosov have introduced in \cite{kl78}
the following application of the Ford-Fulkerson augmenting path
procedure, assuming that a multiflow traverses each edge.
A maximum multiflow unlocks $A\in \K$ if and only if it contains
an \textit{augmenting sequence} $P_{1},x_{1},...,x_{i-1}P_{i}x_{i},....,P_{n}$
of paths $P_{1}$ (an $A$-path), $P_{2},...,P_{n-1}$ ($(A,A^{c})$-paths)
$P_{n}$ (an $A^{c}$-path) and inner nodes 
$x_{1},...,x_{n-1}$ so that $x_{i}\in P_{i},P_{i+1}$ for $i\in {1,...,n-1}$
and $x_{i}$ is located on $P_{i}$ between $x_{i-1}$ and the $A$-end 
of $P_{i}$. 
In the paper, we use the fact that unlocking a member of $\K$
and existence of the alternating sequence are equivalent.
When $\K$ is a K-clutter, there exists a series of switches of $P_{1},...,P_{n}$
in $x_{1},...,x_{n-1}$ that creates a maximum multiflow $f\p$
containing a cross and having $\Theta (f\p)\geq \Theta(f)$.
If $f$ solves the $W$-problem and unlocks $A\in \K$,
switching $P_{1},...,P_{n-1}$ in $x_{1},...,x_{n-2}$ creates a multiflow
$f\p$ with $A$-path $P_{0}\p$ and $A^{c}$-path $P_{1}\p$ having a common node $x_{n-1}$,
so that every switch of $P_{0}\p$ and $P_{1}\p$ in $x_{n-1}$ preserves $\Theta(f\p)=\theta$.

\begin{figure}
{\par \centering
  \resizebox*{0.7\textwidth}{0.2\textheight}{\includegraphics{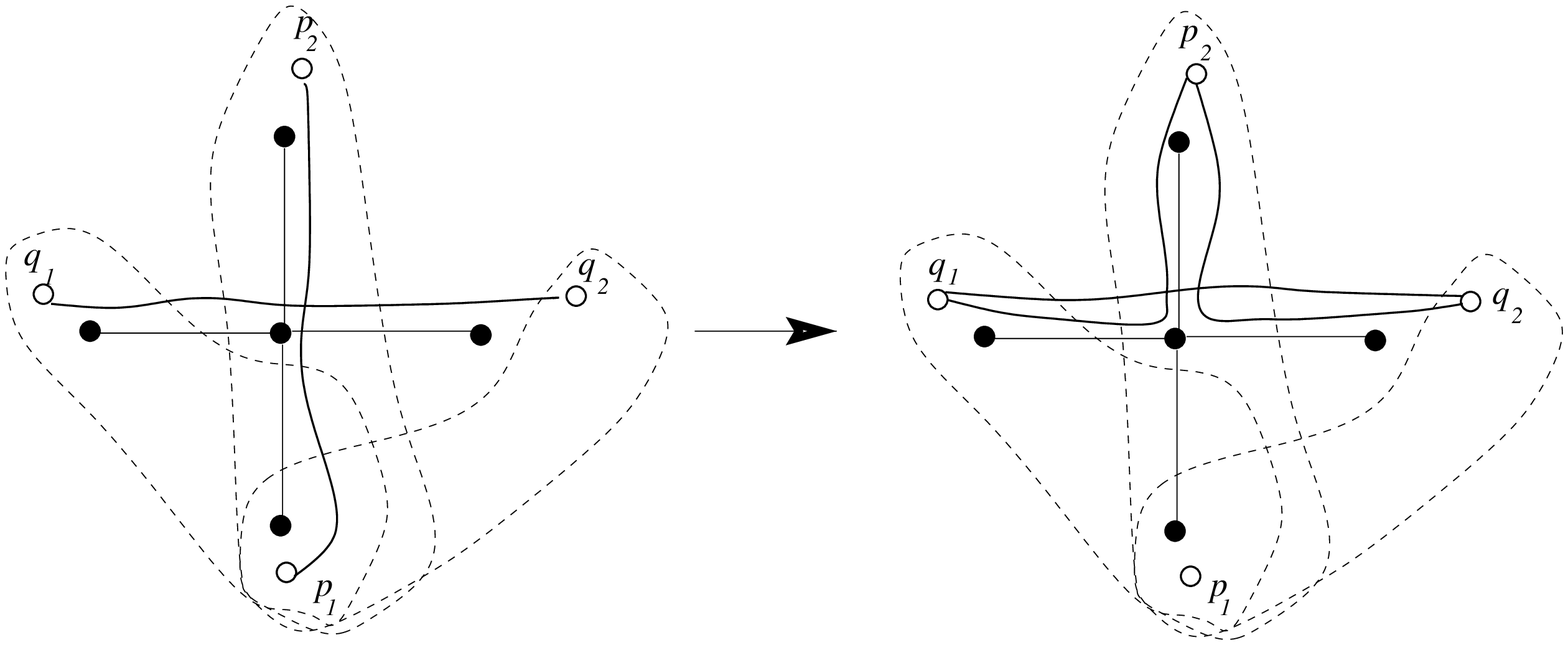}} \par}
\caption{\label{3/2-operation-figure}The $\frac{3}{2}$-operation.}
\end{figure}

Let $P$ and $Q$ be an $A$- and $A^{c}$-paths of a multiflow $h$ with a common inner node so that
$w(P)=w(Q)$ and no switch of $P$ and $Q$ changes $\Theta(h)$. Let us denote the ends of
$P$ and $Q$ by $p_{1},p_{2}$ and $q_{1},q_{2}$ respectively. Let w.l.o.g.
$(p_{1},p_{2}),(p_{1},q_{1}),(p_{1},q_{2})\in W$, $(p_{2},q_{1}),(p_{2},q_{2}),(q_{1},q_{2}),\in S$.
A multiflow transformation that replaces $P$ and $Q$ with three $(p_{2},q_{2})$-,
$(p_{2},q_{2})$- and $(q_{1},q_{2})$-paths of weight
$\frac{w(P)}{2}$ (see Figure \ref{3/2-operation-figure}), is called a
$\frac{3}{2}$\textit{-operation}. It preserves $\Theta(h)$ and
increases $h[S]$ by $\frac{w(P)}{2}$.
  
\section{\label{WP-section}Fractionality of the $W$-problem}
To prove the fractionality conjecture for the $W$-problem,
we show the following:
\begin{theorem}\label{kc-for-WP}In every K-network $\GTK$ there exists a
  simple \WP solution of the smallest size that is half-integer.\end{theorem}
We later use this Theorem to prove the fractionality conjecture for the
$S$-problem.
Let us observe
a K-network $\GTK$ which is a minimal counterexample to Theorem \ref{kc-for-WP}.
We assume that $\GTK$ has \textbf{inner node
  degree $4$}, by the known reduction (see, e.g. \cite{frank}),
  is \textbf{simple} (since atom compression preserves all \WP solutions)
  and is \textbf{minimal} first in fractionality $k$ of the smallest size \WP
  solution,
  and then in $E$ as a set. Then $k=4$, for otherwise we can
  duplicate each edge in $E$ and obtain a network with \WP fractionality $\lceil\frac{k}{2}\rceil$. 
In this section, $f$ denotes \textit{a quarter-integer simple multiflow of the smallest size
solving the \WP} in $\GTK$. For simplicity, we assume that
the paths of $f$ have weight $\quarter$.
Let us denote
\begin{equation}\label{heta}\mbox{\sl $\heta$:=maximum of the \SP among simple multiflows in $\GTK$.}\end{equation}
In the Appendix we prove the max-min theorem for the \WP
in Theorem \ref{wp-maxmin-theorem}, which implies that for every K-network
$\GTK$, $2\theta\GTK\in \mathbb{N}$ and $2\heta\in \mathbb{N}$.
We use these facts in the proof.
\subsection{General flow properties}
Here, we study the behavior of \WP solutions
inside the members of $\K$. The series of properties below directly follows directly from
the results of Lov\~{a}sz, Cherkassky and Lomonosov
described in Section \ref{locking-section}.
\begin{claim}\label{at-least-beta}Let $\GTK$ be a simple K-network, and let $h$ be a simple
  multiflow of fractionality $k$ in it such that $h[A]<\beta(A)$ for some
$A\in \K$. Then there exists a simple multiflow $h\p$ of fractionality $k$
having $\Theta(h\p)\geq \Theta(h)+\half(\beta(A)-h[A])$.\end{claim}
\proof Since $h[A,A^{c}]\leq \lambda(A)$
by definition, and $$h[A]=\half(\sum_{t\in
  A}h[t,t^{c}]-h[A,A^{c}])<\half(\sum_{t\in A}\lambda(t)-\lambda(A))=\beta(A),$$
$\sum_{t\in A}h[t,t^{c}]<\sum_{t\in A}\lambda(t)$. We modify $h$ by adding
paths starting in $t\in A$ until $h[t,t^{c}]=\lambda(t)$ for all $t\in A$.
Since we use edges not saturated by $h$, we obtain a simple multiflow
of fractionality $k$, denoted $h\p$. If $W$- or $S$-paths of total
weight no less than $\beta(A)-h[A]$ were added,
$h\p$ is the required multiflow. Otherwise, some of these paths
are cycles that traverse one terminal from $A$ each.
%We have $\sum_{t\in A}h\p[t,t^{c}]=\sum_{t\in A}\lambda(t)$.
Let us modify $h\p$ into a multiflow without cyclic paths traversing
terminals from $A$ using Cherkassky procedure, and denote the resulting
multiflow by $h\pp$. If $\Theta(h\pp)\geq \Theta(h)+\half(\beta(A)-h[A])$, we are done.
Otherwise, we have $\sum_{t\in A}h\pp[t,t^{c}]=\sum_{t\in A}\lambda(t)$
and $h\pp[A]<\beta(A)$, thus $h\pp[A,A^{c}]>\lambda(A)$ - a
contradiction. \qed
\begin{corollary}\label{smallest-solution-locks-K}Let $\GTK$ be a simple
  K-network, and let $h$ be a simple
  multiflow of the smallest size solving the \WP in $\GTK$.
Then $h$ locks $\K$.\end{corollary}
\proof By Claim \ref{at-least-beta}, $h[A]\geq \beta(A)$ for all $A\in \K$.
If $h$ unlocks some $A\in \K$, i.e. has $h[A]>\beta(A)$, $h$ contains an
augmenting sequence for $A$. Switching paths of this sequence creates a
simple multiflow $h\p$ that has the same size as $h$, solves the \WP and
  allows us to perform a $\frac{3}{2}$-operation,
which preserves $\Theta(h\p)$ but decreases the size of $h\p$ -
a contradiction. \qed

\subsection{Proof of the weak fractionality theorem}
%In this section, we prove the fractionality conjecture for the
%$W$-problem through the following series of claims.
Let us denote by $\GTKp$ a network obtained from $\GTK$ by split-offs in one
or more inner nodes. We denote the \WP maximum in $\GTKp$ by $\theta\p$,
and let $A\p$ and $t\p$ denote a clutter member and a terminal
corresponding to some $A\in \K$ and $t\in T$. We let $g$ denote a simple
half-integer \WP solution of the smallest size in $\GTKp$. $g$ exists
because $\GTK$ is minimal in $E$. Let us denote the value of \eqref{heta}
in $\GTKp$ by $\heta\p$.
Note that
\begin{equation}\label{smaller-eta}\heta\p\leq \heta,\end{equation}
because by Theorem 1 from \cite{va07a}
$f$ solves the \SP in a network obtained from
$\GTK$ by splitting every terminal $t$ into $d(t)$ equivalent terminals
of degree $1$.

For this type of networks
we prove the following series of claims.
\begin{claim}\label{beta-decreases-or-same}Let $\theta\p=\theta-\half$ and
  $\heta-\heta\p\leq 1$. Then
$\sum_{A\p\in \K\p}\beta(A\p)\leq \sum_{A\in \K}\beta(A)$.\end{claim}
\proof Let us assume that $\sum_{A\p\in \K\p}\beta(A\p)>\sum_{A\in \K}\beta(A)$.
As all $\beta(A)$ and $\beta(A\p)$ are integers by definition,
we have $$\theta-\theta\p=\half=\heta-\heta\p+(\sum_{A\in \K}\beta(A)-\sum_{A\p\in
  \K\p}\beta(A\p)),$$
thus 
$$1\geq \heta-\heta\p=\half+\sum_{A\p\in \K\p}\beta(A\p)-\sum_{A\in
  \K}\beta(A)>1,$$ a contradiction. \qed
\begin{corollary}\label{beta-increases-or-same}Let $\theta\p=\theta-\half$ and
  $\heta-\heta\p\leq 1$. Then for all $A\in \K$, $\beta(A\p)\geq
  \beta(A)$.\end{corollary}
\proof Let $\beta(A\p)<\beta(A)$. Then by Claim \ref{at-least-beta},
$g$ can be completed to a half-integer simple flow $g\p$ in $\GTK$
with $\Theta(g\p)=\theta$. Since $|g|=\heta\p+\sum_{A\p\in \K\p}\beta(A\p)<|f|$
by Claim \ref{beta-decreases-or-same} and \eqref{smaller-eta}, we have
$|g\p|\leq |f|$ - a contradiction. \qed
\begin{corollary}\label{same-beta-eta-minus-half}Let $\theta\p=\theta-\half$ and
  $\heta-\heta\p\leq 1$. Then for all $A\in \K$,
  $\beta(A\p)=\beta(A)$ and $\heta-\heta\p=\half$.\end{corollary}
\proof Follows from Claim  \ref{beta-decreases-or-same}
and Corollary \ref{beta-increases-or-same}. \qed

\begin{claim}\label{theta-decreases}$\theta\p\neq \theta$.\end{claim}
\proof Let us assume the contrary. Then for all $A\in \K$,
$\beta(A\p)\geq \beta(A)$, for otherwise by Claim \ref{at-least-beta},
in $\GTK$ $g$ can be modified into a multiflow $g\p$ with
$\Theta(g\p)>\theta$ - a contradiction.
If $\sum_{A\p\in \K\p}\beta(A\p)>\sum_{A\in \K}\beta(A)$,
we have $$\theta-\theta\p=0=\heta-\heta\p+(\sum_{A\p\in \K\p}\beta(A\p)-\sum_{A\in
  \K}\beta(A))>1,$$
a contradiction because $\heta>\heta\p$ (otherwise, $g$ is the solution we
seek).
Then $g[W]=f[W]=\sum_{A\in \K}\beta(A)$ and $\Theta(g)=\Theta(f)$,
resulting in $|g|=|f|$ - a contradiction. \qed

\begin{figure}
  {\par \centering
     \resizebox*{0.9\textwidth}{0.3\textheight}{\includegraphics{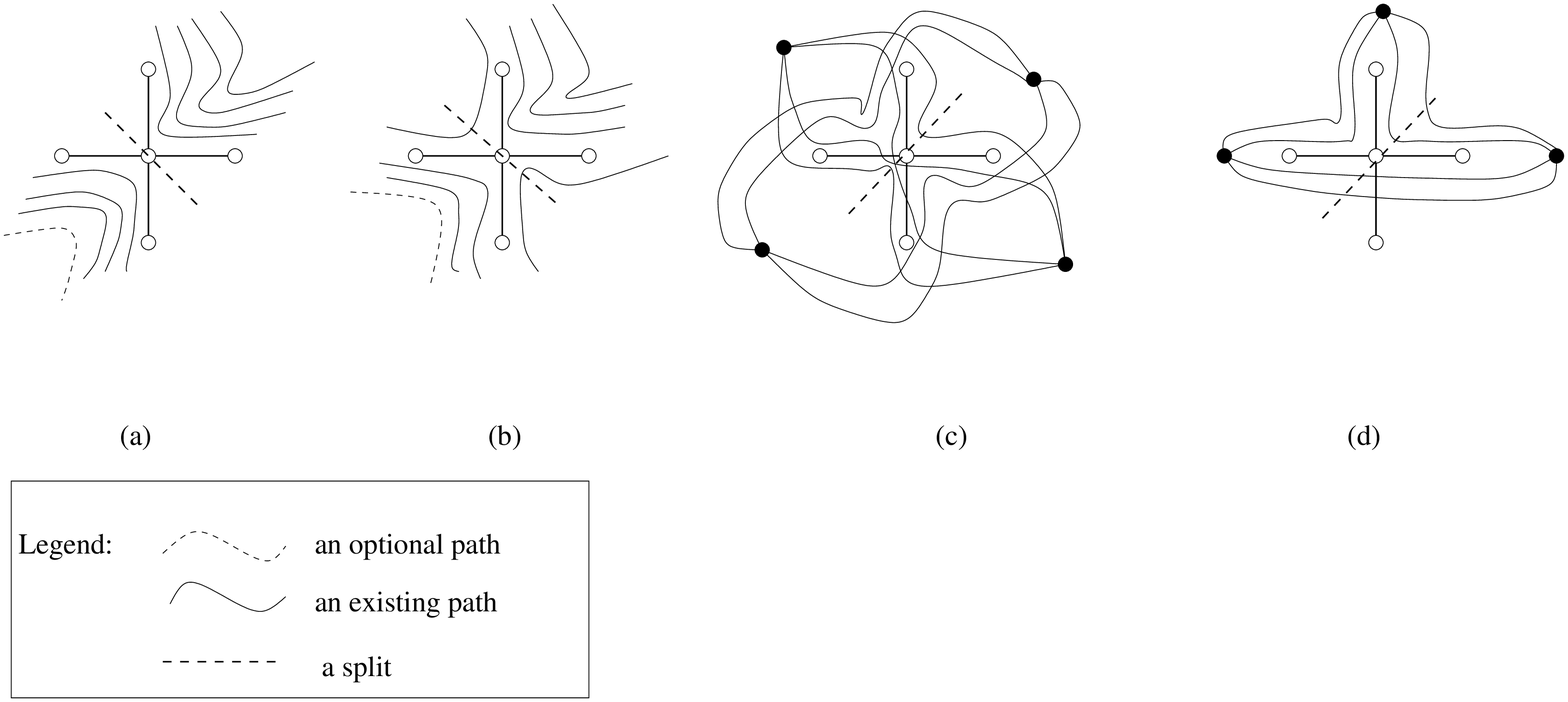}} \par}
   \caption{\label{good-split-figure}Possible switches of $f$ in an inner node.}
  \end{figure}

Let us call two paths traversing the same inner node $x$
\textit{opposite in} $x$ if they do not
traverse the same edge incident to $x$.
\begin{claim}\label{good-split}Let $x\in N\setminus
  T$. Then there exists a
  split of $x$ that decreases
  $\theta$ by no more than $\half$.\end{claim}
\proof Let us assume the contrary. Let the number of paths of $f$ destroyed by a split of $x$
  be $n$. Then the split decreases
  $\Theta(f)$ by at least $1$ by Corollary \ref{theta-half-integer},
  thus $8\geq n\geq 4$. Clearly, $n\neq 7,8$
  for otherwise $x$ admits an $f$-split (see Figure \ref{good-split-figure}(a)).
  Likewise, if $n\in \{5,6\}$, then the switch opposite to the chosen one destroys
  no more than two paths of $f$ (see Figure \ref{good-split-figure}(b)) - a contradiction.
  Therefore, $n=4$, and the paths destroyed by a split
  contribute no more than $1$ to $\Theta(f)$.
  By our assumption, the split decreases $\Theta(f)$ by $1$, and these paths
  are $S$-paths of $f$ with two common ends.
  By our assumption, two of these paths cannot be
  switched so as to comply with the remaining paths traversing $x$.
  If these two paths are opposite,
  we switch one pair so as to comply with the other, and there are
  two options to do so (see Figure \ref{good-split-figure}(c)). The opposite switch affects
  the other $4$ paths of $f$ traversing $x$ and, like above, those paths can traverse $x$
  in two different ways. We then select a common switch and obtain
  a new multiflow $f\p$ that is a common solution in $\GTK$ and admits
  an $f\p$-split in $x$ - a contradiction.
  If the paths in question are not opposite (see Figure \ref{good-split-figure}(d)),
  all the paths of $f$ traversing $x$ end in two terminals. Then there
  exists a switch of paths of $f$ in $x$ allowing an $f$-split - a
  contradiction. \qed

We can now finish the proof of the fractionality theorem for the $W$-problem.
\begin{atheorem}{\ref{kc-for-WP}}Let $\GTK$ be a K-network. Then in $\GTK$ there
  exists a simple half-integer \WP solution of the smallest size.\end{atheorem}
\proof Let $\GTKp$ be the network with $\theta\p=\theta-\half$ and
$\heta-\heta\p\leq 1$, obtained from $\GTK$ by the maximum number of
split-offs in inner nodes. At least one such network exists because of
Claim \ref{good-split}. By Claim \ref{theta-decreases} and Corollary
\ref{same-beta-eta-minus-half}, $\beta(A\p)=\beta(A)$ for all $A\in \K$.
Then $\heta-\heta\p=\half$.

Let $g$ denote a simple \WP solution of the
smallest size in $\GTK$. 
Since $|g|=|f|-\half$, $g$ is not maximum and
we can add a half-integer zero path $P$ to $g$ with an end in $t\in A$.
We select $g$ so that $P$ is the longest w.r.t. number of edges.
Let $P$ traverse edge $(t,x)$. Then a path $Q\in g$ opposite to $P$
in $x$ has no end in $t$ (otherwise, switching $P$ and $Q$ prolongs $P$).
\begin{figure}
  {\par \centering
     \resizebox*{0.6\textwidth}{0.2\textheight}{\includegraphics{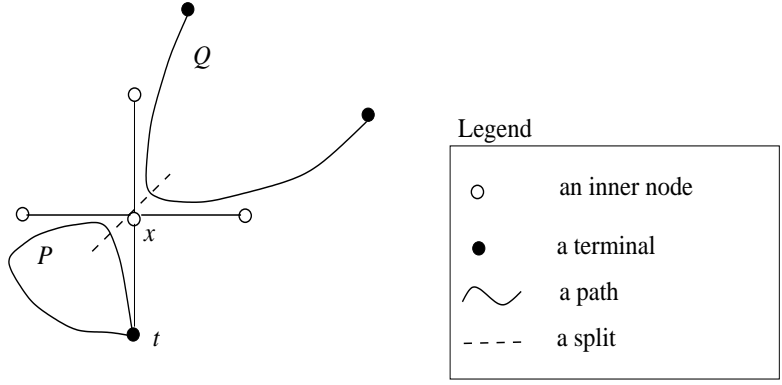}} \par}
   \caption{\label{zero-cycle-figure}$\theta$-preserving split of an inner node.}
\end{figure}

Switching of $P$ and $Q$ in $x$ cannot increase
$g[S]$ for then the resulting half-integer flow $g\p$ has
$\Theta(g\p)=\theta$ and $|g\p|\leq |f|$. Likewise, switching $P$ and $Q$ so
as to allow a $g$-split in $x$ cannot increase $\Theta(g)$, for
otherwise we obtain a network $\GTKpp$ with $\theta\pp\geq
\theta-\frac{1}{4}$ - a contradiction to Claim \ref{theta-decreases}.
Therefore, $Q$ is a $t^{c}$-path and an $S$-path. Switching $P$ and $Q$
in $x$ so as to allow a $g$-split of $x$ produces two $W$-paths (see Figure \ref{zero-cycle-figure}).
We switch $P$ and $Q$ in this way, obtain a new multiflow $g\pp$ and
a network denoted $\GTKpp$. Then $\theta\pp=\theta-\half$ and
$\heta\pp\geq \heta\p-\half=\heta-1$ while $\GTKpp$ contains less inner nodes
than $\GTKp$, contrary to our choice. \qed

\section{Fractionality of the \SP}
We use Theorem \ref{kc-for-WP} to show that the
fractionality conjecture for the $S$-problem
 holds.
Let us select a K-network $\GTK$ which is a counterexample to the conjecture,
$$\mbox{\sl minimal in fractionality $k$ and $\alpha:=\frac{\sum_{t\in T}|N(t)|}{|T|}$.}$$
Like in Section \ref{WP-section}, we can assume that $k=4$.

\begin{claim}\label{alpha=1}$\alpha=1$\end{claim}
\proof Let us assume the contrary and select
$t\in T$ with $|N(t)|\geq 2$. Let $g$ be a quarter-integer
common solution to the $W$- and $S$-problems in $\GTK$.
Let us suppose first that no path of $g$ has an end in $t$.
We turn $t$ into an inner node, adding 
a new terminal $t\p\sim t$ and an edge $(t,t\p)$ if $d(t)$ is odd.
In the resulting network $\GTKp$,
$\eta\p:=\eta\GTKp=\eta$ because the reverse operation does not decrease $\eta\p$.
Let us suppose now that $g$ contains paths with an end in $t$.
Let $w_{g}(t)$ denote the
total weight of $g$'s paths beginning in $t$.
Then $w_{g}(t)\leq \frac{3}{4} d(t)$, for otherwise there exists an edge $(t,x)$
traversed by four paths of weight $\quarter$ with an end in $t$.
We replace $(t,x)$ with a new edge $(t\p,x)$, where $t\p\sim t$
is a new terminal, and turn $t$ into an inner node.
We also add
enough $(t,t\p)$-edges to allow the paths of $g$ with an end in $t$
to end in $t\p$ instead and the degree of $t$ to be even.
In the resulting network $\GTKp$, $\alpha\p<\alpha$ and
$\eta\p=\eta$ because the reverse operation does not decrease $\eta\p$. \qed

\begin{theorem}\label{kc-for-SP}Every K-network $\GTK$ admits a %add ls reasoning!!!
  half-integer least-size \WP solution $f$ that also solves the $S$-problem.\end{theorem}
\proof Let $\GTK$ be a K-network $\GTK$.
By Claim \ref{alpha=1}, we can transform $\GTK$ into a K-network
$\GTKp$ with $\alpha=1$, $\eta\p=\eta$ and $\theta\p=\theta$.
Moreover, every \SP or \WP solution in $\GTKp$ remains such in $\GTK$
after the reverse transformation.
By Theorem \ref{kc-for-WP}, $\GTKp$ admits a simple half-integer \WP
solution of the smallest size, denoted $f\p$.
By Theorem 1 of \cite{va07a}, $f\p$ solves the \SP in $\GTKp$.
Then the multiflow $f$ in $\GTK$, obtained from $f\p$, solves
both $W$- and $S$-problems. \qed
\begin{corollary}In a general, not necessarily Eulerian, network $(G,T)$
where the anticlique clutter of $(T,S)$ is a K-clutter, both \WP and \SP
have fractionality $4$. \qed\end{corollary}

\section{Acknowledgments}
The author expresses her deepest gratitude to
Prof. Eyal S. Shimony for the help with this manuscript
and the Lynn and William Fraenkel Center for Computer Science
for partially supporting this work.

\section{Appendix: combinatorial max-min for the $W$-problem}
Let $\E=\{\alpha,\beta,...\}$ be a partition of $T$ such that
for each $\alpha\in \E$ any $t\p,t\pp\in \alpha$ are equivalent (an
\textit{equi-partition}).
We call
$\x=(X_{\alpha}:\: \alpha\in \E)$ is an \textit{expansion}
if $X_{\alpha}\cap T=\alpha$, $\alpha\in \E$. 
Taking members of $\x$ as terminals and an induced clutter, we obtain a new
network with a graph $\Gx$, terminals $\x$ and a clutter $\kx$ on $\x$
($\kx$ is a K-clutter if $\K$ is a K-clutter).
For $X_{\alpha},X_{\beta}\in \x$,
we call $(X_{\alpha},X_{\beta})$ \textit{strong} or \textit{weak} if for every $s\in \alpha$
and $t\in \beta$,
$(s,t)\in S$ or $(s,t)\in W$ respectively. Likewise, $X_{\alpha}\sim X_{\beta}$ if for
every pair of terminals $s\in \alpha$ and $t\in \beta$, $s\sim t$.
An $\mathcal{X}$\textit{-path} in
$G$ is an $(x,y)$-path with $x,y$ lying in distinct members of
$\mathcal{X}$.
An $\mathcal{X}$\textit{-flow} is a flow in the network $\GTKx$
consisting of $\mathcal{X}$-paths.
The \SP and the \WP in $(\Gx,\x,\kx)$ are defined in the same way as for
$\GTK$, and their maxima are denoted by $\etax$ and $\thetax$ respectively.

We define a \textit{partial order} on expansions as follows.
Let $\E$ and $\F$ be equi-partitions of $T$ and let
$\x=(X_{\alpha}:\: \alpha\in \E)$ and
$\y=(Y_{\alpha}:\: \alpha\in \F)$ be expansions.
Then $\mathcal{X}\preceq \mathcal{Y}$ if for every $X\in \x$
there exists $Y\in \y$ so that $X\subset Y$.
Note that for every $\mathcal{X}\preceq \mathcal{Y}$, every $\mathcal{X}$-flow
is also a $\mathcal{Y}$-flow (but the converse may be not true).
Since for $\mathcal{X}\preceq \mathcal{Y}$ any $\mathcal{X}$-flow
is also a $\mathcal{Y}$-flow, $\thetay\geq \theta_{\mathcal{X}}$.
Since $T$-flow is also an $\mathcal{X}$-flow,
$\theta_{\mathcal{X}}\geq \theta$. 
$\mathcal{X}$ is called \textit{critical}
if $\thetay>\thetax$ for every $\y\succ \x$. A critical $\x$ with
$\theta_{\x}=\theta$ is called a \textit{dual solution}.
The triangle theorem (\cite{l85})
ensures that:
\begin{equation}\label{max-x-solution}\mbox{\textsl{there exists a maximum $\mathcal{X}$-flow
$h$ such that}
$\Theta_{\mathcal{X}}(h)=\theta_{\mathcal{X}}$.}\end{equation}
We limit ourselves to networks $\GTK$ with simple $\K$.
The results of this section that hold for simple clutters
hold for general networks as well, because compressing a non-trivial
atom into one terminal does not change $\theta$ by 
triangle theorem from \cite{l85} and metric properties of a K-clutter.
For a K-network with simple $\K$, every subset in an expansion $\x$ contains
exactly one terminal; $X_{t}$ denotes a member of $\x$ containing $t\in T$.
Then \eqref{max-x-solution} implies that for a maximum $X$-flow $h$ (even
when $\x=T$):
\begin{equation}\label{theta-with-size}\Theta_{\mathcal{X}}(h)=|h| - \half
  h[W].\end{equation}
We aim to prove the following max-min theorem for the fractional $W$-problem.
\begin{theorem}\label{wp-maxmin-theorem}
  In a K-network $\GTK$:
\begin{equation}\label{wp-maxmin-equation2}\mbox{\sl $\mathrm{max}_{f} \Theta(f)=\mathrm{min}_{\x}( \frac{1}{2}\sum _{t\in
  T}d(X_{t})-\frac{1}{2}\sum _{A\in \kx}\beta(A))$.}\end{equation}
The maximum is taken over the fractional multiflows in $\GTK$, and the
minimum is taken over all expansions in $\GTK$.
Moreover, \eqref{wp-maxmin-equation2} holds as equality for every dual
solution
$\x$.\end{theorem}
To prove this theorem, we state the following inequality for
an expansion $\mathcal{X}$ and a $T$-flow $f$:
\begin{equation}\label{max-min}\Theta (f)\begin{array}{c}
 (a)\\
 \leq \\
 \, \end{array}
\theta \begin{array}{c}
 (b)\\
 \leq \\
 \, \end{array}
\Theta _{\mathcal{X}}(h)
\begin{array}{c}
 (c)\\
 \leq \\
 \, \end{array}
\frac{1}{2}\sum _{t\in T}d(X_{t})-\frac{1}{2}\sum _{A\in \kx}\beta(A)
\end{equation}
We aim to show that \eqref{max-min} holds as
inequality for every expansion and as equality for every critical expansion.
\eqref{max-min}(a) follows directly from the definition of $\theta $. 
\eqref{max-min}(b) holds because
$f$ is also an $\mathcal{X}$-flow. \eqref{max-min}(c) holds because there exists a maximum
$\mathcal{X}$-flow $h$ that solves the \WP in $\x$.
For such $h$ the minimum of $\sum _{A\in \kx}h[A]$ is
achieved when all $A\in \kx$ are locked by $h$, i.e.
$\sum _{A\in \kx}h[A]\leq \sum _{A\in \kx}\beta(A)$
and $|h| =\frac{1}{2}\sum _{t\in T}\lambda (X_{t})$
by the \Lovasz-Cherkassky theorem (\cite{lo76,ch77}).
We need the following two claims to show that \eqref{max-min}(c) is an
equality.
\begin{claim}\label{saturation}Let $\GTK$ be a simple K-network, and
  let $\mathcal{X}$ be a dual solution in it.
  A maximum fractional $\mathcal{X}$-flow $h$ that
satisfies $\Theta _{\mathcal{X}}(h)=\thetax$
(that is, solves the \WP in $\GTKx$) locks $X_{t}$ for all $t\in T$.\end{claim}
\proof First, let us show that $h$ saturates every $(X_{t},\overline{X_{t}})$-edge.
Let $e$ be an $(x,y)$-edge with $x\in X_{t}$ and $y\in \overline{X_{t}}$. Let
$\mathcal{Y}\succ \mathcal{X}$ be an expansion where $Y_{s}=X_{s}$ for
terminal $s\neq t$ and $Y_{t}=X_{t}\cup \{y\}$. Since
$\mathcal{X}$ is critical, $\thetay>\thetax$
and there exists a $\mathcal{Y}$-flow $g$ such that $\Theta _{\mathcal{Y}}(g)>\thetax$.
Let us
denote the unused capacity of $e$ by $\varepsilon $ and let $\delta =g[y,\cup _{s\neq t}X_{s}]$.
%If $\varepsilon \geq \delta $, we can modify $g$ by turning $x$
%into a source of $(y,\cup _{s\neq t}X_{s})$-paths instead of $y$.
%Then $g$ becomes an $\mathcal{X}$-flow , a contradiction. 
Clearly, $\varepsilon <\delta $. We turn $g$ into an $\mathcal{X}$-flow
by prolonging all its paths starting in $y$ to $x$ instead through the edge
$e$. Let $g\p$ be the functions on $\x$-paths thus obtained; $g\p$  
does not satisfy the capacity constraint on $(x,y)$.
Then there exists $0<\alpha<1$ such that $h\p=(1-\alpha )h+\alpha g\p$
is an $\x$-flow.
%we build a new fractional $\mathcal{X}$-flow as a positive linear
%combination of $h$ and $g$: $h\p=(1-\alpha )h+\alpha g$ for some
%$0<\alpha <1$ such that $(1-\alpha )\varepsilon \geq \alpha \delta $
%(which means that $\alpha \leq \frac{\varepsilon }{\varepsilon +\delta }$).
$h\p$ satisfies all capacity
constraints and has $\Theta _{\mathcal{X}}(h\p)\geq (1-\alpha )\Theta _{\mathcal{X}}(h)+\alpha \Theta _{\mathcal{Y}}(g)>\thetax$,
contradicting the definition of $\mathcal{X}$.

Let us assume now that a $(p,q)$-path $P$ of $h$, $p\in X_{t}$, contains
two $(X_{t},\overline{X_{t}})$-edges, $e_{1}=(x_{1},y_{1})$ and
$e=(x_{2},y_{2})$ where $x_{1},x_{2}\in X_{t},y_{1},y_{2}\in \overline{X_{t}}$
and $y_{1},x_{1},x_{2},y_{2}$ appear on $P$ in this order. Then
by replacing $P$ with $x_{2}Pq$ we obtain an $\mathcal{X}$-flow
$g$ for which $\Theta _{\mathcal{X}}(g)=\thetax$ and
the edge $(x_{1},y_{1})$ is not saturated by $g$, a contradiction.
\qed

\begin{claim}\label{locking-clutter} Let $\GTK$ be a simple K-network, and
  let $\mathcal{X}$ be a dual solution. A maximum fractional $\mathcal{X}$-flow $h$
would then satisfy $\Theta _{\mathcal{X}}(h)=\thetax$
iff every $A\in \kx$ is locked by $h$.\end{claim}
\proof The ``if'' direction is trivial.
Let $h$ be a maximum $\x$-flow with $\Theta _{\mathcal{X}}(h)=\thetax$
that locks every member of $\kx$.
Because of Claim \ref{saturation} and the simplicity of $\kx$, we get
$\Theta(h)=\frac{1}{2}\sum _{X\in \x}d(X)-\frac{1}{2}\sum _{A\in
  \kx}\beta _{A}$
and thus $\Theta(h)\geq \thetax$ by \eqref{max-min}(c).

For the ``only if'' direction, assume that $h$ is a maximum $\x$-flow that has
$\Theta_{\mathcal{X}}(h)=\thetax$ and unlocks $A\in \kx$.
Let $A^{c}$ in the context of $\kx$ denote the members of $\x$ that do not lie in $A$.  
Then $h$ contains an augmenting
sequence  $P_{0},x_{0},...,x_{m-1},P_{m}$,
where $P_{0}$ is an $A$-path,
$P_{m}$ is an $A^{c}$-path,
and each one of $P_{1},...,P_{m-1}$ is an $(A,A^{c})$-path.
We can choose $h$ so that $m=1$.
Let $P_{0}$ and $P_{1}$ be $(s\p,t\p)$- and $(q\p,r\p)$-paths
with weights $\alpha $ and $\beta $ respectively where $s\p\in X_{s}$,
$t\p\in X_{t}$, $q\p\in X_{q}$ and $r\p\in X_{r}$.
Since a switch of $P_{0}$ and $P_{1}$ in $x_{0}$ cannot increase
$\Theta(h)$, we can assume that w.l.o.g. $(X_{q},X_{r})$, $(X_{t},X_{r})$
and $(X_{t},X_{q})$ are $S$-pairs while $(X_{s},X_{q})$ and $(X_{s},X_{r})$
are $W$-pairs by the simplicity of $\kx$.

\begin{figure}
{\par \centering
  \resizebox*{0.3\textwidth}{0.15\textheight}{\includegraphics{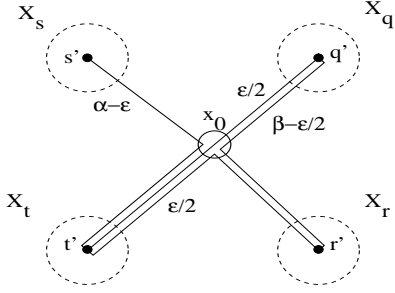}} \par}
\caption{\label{3/2-operation}The fractional $\frac{3}{2}$-operation.}
\end{figure}

We construct a new flow $f$ from $h$ by replacing $P_{0}$ and $P_{1}$
with $(t\p,r\p)$, $(t\p,q\p)$, $(q\p,r\p)$ and $(s\p,t\p)$-paths of weights
$\frac{\varepsilon }{2}$, $\frac{\varepsilon }{2}$, $\beta -\frac{\varepsilon }{2}$
and $\alpha -\varepsilon $ respectively (this is the $\frac{3}{2}$\textit{-operation},
see Figure \ref{3/2-operation}). It follows that $|f| =|h| -\frac{\varepsilon }{2}$
and $f[W]= h[W]-\varepsilon$
since $(X_{q},X_{t}),(X_{q},X_{r}),(X_{r},X_{t})\in S$
            and $\Theta _{\mathcal{X}}(f)=\Theta _{\mathcal{X}}(h)$.

The subpath $s\p P_{0}x_{0}$ does not have common nodes with any other
$\mathcal{X}$-path $Q$ whose ends do not lie in $X_{s}\cup X_{t}$.
If it were so, then the above $\frac{3}{2}$-operation could be applied
to both $P_{0},P_{1}$ and $P_{0},Q$ and a flow $f\p$ with $|f\p| =|h| -\frac{\varepsilon}{2}$
and $f\p[W]=h[W]-2\varepsilon$
could be created, which contradicts the maximality of $\Theta _{\mathcal{X}}(h)$.
Therefore, there exists an edge $(s\p,x)$ of $s\p Lv$ which is not
saturated by $f$ - a contradiction to Claim \ref{saturation}.
\qed

Theorem \ref{wp-maxmin-theorem}
follows from Claims \ref{saturation} and \ref{locking-clutter}.
\begin{corollary}\label{theta-half-integer}$2\theta\GTK\in
  \mathbb{N}.$\end{corollary}
\proof Let $\x$ be an expansion that achieves equality in Theorem
\ref{wp-maxmin-theorem} for $\GTK$. Then
$\theta\GTK=\half\sum_{X\in \x}d(X)-\half\sum_{A\in \kx}\beta(A)$,
while $\sum_{X\in \x}d(X)$ is always even in an Eulerian network and
every $\beta(A)$ is an integer by definition. Thus, a split of an inner node
  in $\GTK$ decreases $\theta$ by $\frac{k}{2}$, $k\in
  \mathbb{N}\cup \{0\}$. \qed
\begin{corollary}\label{eta-half-integer}
  Let $\GTK$ be a simple K-network and let $h$ be a simple \WP solution in $\GTK$
 with $\sum_{A\in\K}h[A]=\sum_{A\in \K}\beta(A)$.
  Then $2h[S]\in \mathbb{N}$.\end{corollary}
\proof $2h[S]$ is an integer because  $\theta=h[S]+\half h[W]=h[S]+\half \sum_{A\in
  \K}\beta(A)$ and $\theta$ is half-integer by Corollary \ref{theta-half-integer}. \qed

\renewcommand{\baselinestretch}{1}
\bibliographystyle{plain}

\renewcommand{\baselinestretch}{1.5}
\newpage
\begin{center}\large \bf Keywords\end{center}

Path packing, multiflow, fractionality

\newpage
\begin{center}\large \bf Contact author\end{center}

Natalia Vanetik

Department of Computer Science, Ben-Gurion University, Israel.

\textbf{E-mail address}: \textit{orlovn@cs.bgu.ac.il}

\textbf{Fax}: +972-8-6477650

\textbf{Phone}: +972-8-6477866

\textbf{Address}:

\hspace{2cm}N. Vanetik

\hspace{2cm}Department of Computer Science

\hspace{2cm}Ben Gurion University of the Negev

\hspace{2cm}P.O.B 653 Be'er Sheva 84105

\hspace{2cm}Israel

\newpage
\begin{center}\large \bf Footnotes\end{center}

\textbf{Author affiliation:}

N. Vanetik, Department of Computer Science, Ben-Gurion University,
Israel, \textit{orlovn@cs.bgu.ac.il}

\end{document}